\documentclass[12pt]{article}
\usepackage{graphicx}
\usepackage[cp1251]{inputenc}
\usepackage{epsfig}
\usepackage[english]{babel}

\date{}

\begin{document}
\setcounter{page}{1}
\pagestyle{plain}

\title{\bf{Integer quantum Hall effect in AAA-stacked trilayer graphene}}
\author{Yawar Mohammadi\thanks{Electronic address:
y.mohammadi@cfu.ac.ir}} \maketitle{\centerline{Department of
Physics, Farhangian University, Tehran,
Iran}

\begin{abstract}

We obtain an analytical expression for the Hall conductivity in AAA-stacked trilayer graphene (TLG) within the linear response Kubo formalism. Interestingly, we find that the quantization of the Hall conductivity is severely dependent on the value of the chemical potential and the magnetic field. The origin of this unusual quantization, attributed to the Landau level spectrum of AAA-stacked TLG, is explained in detail. We show that is a characteristic of the multilayer graphene structures with AA stacking order.

\end{abstract}


\vspace{0.5cm}

\section{Introduction}
\label{sec01}

Graphene, an isolated layer of graphite, due its chiral 
low-energy spectrum dominated by a massless Dirac-like equation, shows many intriguing properties~\cite{Castro Neto1,Peres1}. Further, properties of multilayer
graphene materials, as have been found in the recent researches~\cite{Partoens1,Koshino1,Nilsson1,Avetisyan1,Koshino2,Das Sarma1,Jung1,de Andres1,Munoz1},
are dependent on their stacking order and the number of
layers. The most of
these studies have been devoted to the multilayer graphene
structures with Bernal (ABA) and rhombohedral (ABC) stacking order. Ab initio density functional
theory calculations~\cite{Aoki1,Andres1}, and also STM and TEM imaging~\cite{Lauffer1,Liu1} of multilayer graphene show that a new stable stacking order of few-layer graphene (with
AA stacking order) is also possible. In multilayer graphene structures with AA
stacking order, each sublattice in the upper layers is located directly
above the same one in the lower layers. Consequently, AA-stacked multilayer graphene structures have special low energy band structures which
are composed of electron-doped, hole-doped and even undoped
SLG-like band structures~\cite{Borysiuk1,Ando1}.
For some properties, these SLG-like bands behave like decoupled bands leading to many
attractive properties which are different form that of SLG and
also have not been observed in the other graphene-based
materials~\cite{Hsu1,Prada1,Tabert1,Brey1,Mohammadi1}.

Integer quantum Hall effect is one of the most intriguing properties of two dimensional electron gas systems in which the non-diagonal (Hall) conductivity in a strong magnetic field is quantized in integer multiple of $e^{2}/h$~\cite{Klitzing1}. This effect has been observed in graphene with a different quantization, originating from  its chiral massless Dirac-like spectrum~\cite{Gusynin1,Novoselov1}. Furthermore, recently several groups have studied the integer quantum Hall effect in multilayer graphene with different stacking order~\cite{McCann1,Koshino3,Henriksen1,Min1,Min2,Hsu2}. These studies revealed
that the quantization of the Hall conductivity in these
materials not only are dependent on the number of layers and the
stacking order but also have unusual futures which have not been observed in graphene. Motivated by these facts, we study integer quantum Hall effect in AAA-stacked TLG, with some generalization to multilayer graphene structures with AA stacking order. We also comment on one of the main results of Ref. \cite{Hsu2} in which authors have investigated integer quantum Hall effect in AA-stacked bilayer graphene.

The rest of this paper is organized as follows. In the section II,
first we introduce the tight-binding Hamiltonian describing the
low energy quasiparticle excitation in AAA-stacked TLG. Then we
obtain the Landau levels spectrum and the wavefunctions of the AAA-stacked TLG. In the next subsection, using Kubo formula, we obtain an analytical expression for the Hall conductivity in AAA-stacked TLG. In the section III, we investigate the chemical potential and the magnetic filed dependence of the Hall conductivity, and determine its quantization. Finally,
we end our paper by summary and conclusions in the section IV.

\section{Model and Formalism}
\label{sec02}

In this section first we introduce the Hamiltonian of AAA-stacked TLG inexposed to a magnetic field applied perpendicular to its planes. Then we calculate the corresponding eigenvalues and eigenfunction. In the next subsection we obtain an analytical expression for the Hall conductivity of AAA-stacked TLG within the linear response Kubo formalism.


\subsection{Hamiltonian, Landau levels spectrum and Wavefunctions}
\label{subsec01}

In an AAA-stacked TLG lattice which is composed of three SLG, each
sublattice in the upper layers is located directly above the same one
in the lower layers. The unit cell of an AAA-stacked TLG consists
of 6 inequivalent Carbon atoms, two atoms for every layer. AAA-stacked TLG, similar to SLG, has a two dimensional hexagonal
Brilloin zone and its low-energy excitations occur near Dirac
points ($\mathbf{K}$ and $\mathbf{K}^{'}$). Its low-energy Hamiltonian is given by \cite{Mohammadi2}
\begin{eqnarray}\label{eq01}
\hat{H}=\left(
\begin{array}{ccc}
\hat{H}_{SL} & t_{\perp}\hat{1} & \hat{0} \\
t_{\perp}\hat{1} & \hat{H}_{SL} & t_{\perp}\hat{1} \\
\hat{0} & t_{\perp}\hat{1} & \hat{H}_{SL} \\
\end{array}
\right),
\end{eqnarray}
where $t_{\perp}$ is the inter-layer nearest neighbor hopping energy, $\hat{0}$ and $\hat{1}$ are zero and unit $2\times2$ matrices, and $\hat{H}_{SL}$ is the low-energy Hamiltonian of SLG given by\cite{Castro Neto1}
\begin{eqnarray}\label{eq02}
\hat{H}_{SL}= \left(
\begin{array}{cc}
0 & v_{F}(p_{x}\pm ip_{y}) \\
v_{F}(p_{x}\mp ip_{y}) & 0 \\
\end{array}
\right),
\end{eqnarray}
where upper(lower) sign denotes $\mathbf{K}$($\mathbf{K}^{'}$) valley and $v_{F}=3ta/2\hbar$ with $t$ being the intra-layer hopping energy and $a$ the intra-layer carbon-carbon distance. In the presence of a magnetic field, the Hamiltonian is obtained by Peierls substitution, $\mathbf{p}\rightarrow \mathbf{p}+e\mathbf{A}$ where $\mathbf{A}$ is the vector potential and $-e$ is the charge of an electron. We use the Landau gauge $\mathbf{A}=(0,Bx,0)$ for a constant magnetic field $\mathbf{B}=(0,0,B)$ applied perpendicular to the AAA-stacked TLG plane. We use a unitary transformation\cite{Mohammadi2}
\begin{eqnarray}\label{eq03}
\hat{U}=\frac{1}{2\sqrt{2}}\left(
\begin{array}{ccc}
\sqrt{2}\hat{1} &  2\hat{1}    &   \sqrt{2}\hat{1}   \\
2\hat{1}        &  \hat{0}     &     -2\hat{1}       \\
\sqrt{2}\hat{1} &  -2\hat{1}   &   \sqrt{2}\hat{1}   \\
\end{array}
\right),
\end{eqnarray}
where $\hat{0}$ and $\hat{1}$ are zero and unit $2\times2$ matrices, to block-diagonalize the obtained Hamiltonian. Then one can easily obtain the corresponding eigenvalues
\begin{eqnarray}\label{eq04}
\epsilon_{n}^{\lambda,\nu}=\lambda\sqrt{n}\hbar \omega_{c}+\nu \sqrt{2}t_{\perp},~~~~n=0,1,2,3,...
\end{eqnarray}
where $\lambda=+,-$, $\nu=+,0,-$, and $ \omega_{c}=v_{F}\sqrt{2eB/\hbar}$. Obviously, the Landau levels of AAA-stacked TLG are three copies of the Landau levels of SLG, two off them are shifted up and down by $\sqrt{2}t_{\bot}$ with respect to zero energy. The corresponding eigenfunctions are
\begin{equation}\label{eq05}
  \hat{\Psi}_{n}^{\lambda,+}(\bar{x})=\left(
                           \begin{array}{c}
                             \hat{\psi}_{n}^{\lambda}(x\bar{})\\
                             \hat{0} \\
                            \hat{0} \\
                          \end{array}
                         \right)
                         \textmd{,}~~~\Psi_{n}^{\lambda,0}(\bar{x})=\left(
                           \begin{array}{c}
                             \hat{0} \\
                             \hat{\psi}_{n}^{\lambda}(\bar{x}) \\
                             \hat{0} \\
                           \end{array}
                         \right)
                         \textmd{and,}~~~\Psi_{n}^{\lambda,-}(\bar{x})=\left(
                           \begin{array}{c}
                             \hat{0} \\
                             \hat{0} \\
                             \hat{\psi}_{n}^{\lambda}(\bar{x}) \\
                           \end{array}
                         \right)
\end{equation}
for $\nu=+$, $\nu=0$, and $\nu=-$, respectively, in which
\begin{equation}\label{eq06}
 \hat{\psi}_{n}^{\lambda}(\bar{x})=\frac{e^{ik_{y}y}}{\sqrt{L_{y}}}\left(
                                                               \begin{array}{c}
                                                                 -i\lambda a_{n} \varphi_{n-1}(\bar{x})\\
                                                                 b_{n} \varphi_{n}(\bar{x})\\
                                                               \end{array}
                                                             \right),
\end{equation}
where $a_{n}=\sqrt{(1-\delta_{ n,0})/2}$, $ b_{n}=\sqrt{(1+\delta_{ n,0})/2}$ with $\delta$ being the Kronecker delta function, $\bar{x}=x-x_{0}$, and $x_{0}=k_{y}l_{B}^{2}$ where $l_{B}=\sqrt{\hbar}/eB$ is the magnetic length. Moreover, $L_{y}$ is the length of the AAA-stacked TLG along the $y$ direction and $ \varphi_{n}(x)=e^{-x^{2}/2}H_{n}(x)/\sqrt{2^{n}n!\sqrt{\pi}l_{B}}$ are the harmonic oscillator functions. One can obtain the wavefunctions for the $K^{'}$ valley from Eq. \ref{eq06} by interchanging $a_{n}\varphi_{n-1}(\bar{x})$ and $b_{n}\varphi_{n}(\bar{x})$, and the corresponding eigenvalues from Eq. \ref{eq04} with replacing $\lambda$ by $-\lambda$. In the above calculation we supposed $B>0$, namely the magnetic field is in the z direction. It is easy to show that for $B<0$ the eigenvalues and eigenfunctions for each Dirac point is equal to those of other Dirac point for $B>0$. In the next subsection we utilize these results to obtain a general analytical expression for the Hall conductivity in AAA-stacked TLG.

\subsection{Hall conductivity}
\label{subsec02}

In this paper we interest in the Hall conductivity in the linear response regime. Using the Kubo formula, the Hall conductivity can be written as\cite{Krstajic1,Tahir1}
\begin{equation}\label{eq07}
 \sigma_{xy}=g_{s}g_{v}\frac{i\hbar e^{2}}{L_{x}L_{y}}\sum_{\zeta,\zeta^{'}}[f(\epsilon_{\zeta})-f(\epsilon_{\zeta^{'}})]\frac{\langle\zeta|v_{x}|\zeta^{'}\rangle\langle\zeta^{'}|v_{y}|\zeta\rangle}
{(\epsilon_{\zeta}-\epsilon_{\zeta^{'}})^{2}},
\end{equation}
where $g_{s}=2$ and $g_{v}=2$ are due to two fold spin and valley degeneracy, $f(\epsilon_{\zeta})=1/(1+e^{(\epsilon_{\zeta}-\mu)/k_{B}T})$ is the Fermi Dirac distribution function with $\mu$ being the chemical potential, $|\zeta\rangle=|n,k_{y},\lambda,\nu\rangle$ are the eigenfunctions of the AAA-stacked TLG written in the Dirac notation and $\epsilon_{\zeta}$ are the corresponding eigenvalues (Landau levels spectrum). The $x$ and $y$ components of the velocity operator can be obtained from the Hamiltonian via $v_{x}=\frac{\partial H}{\partial p_{x}}$ and $v_{y}=\frac{\partial H}{\partial p_{y}}$. After evaluating
\begin{equation}\label{eq08}
 \langle\zeta|v_{x}|\zeta^{'}\rangle=\frac{v_{F}}{2}(\lambda^{'}a_{n^{'}}b_{n}\delta_{n^{'},n+1}+\lambda a_{n}b_{n^{'}}\delta_{n^{'},n-1})\delta_{\nu,\nu^{'}}\delta_{k_{y}k_{y}^{'}},
\end{equation}
and
\begin{equation}\label{eq09}
 \langle\zeta^{'}|v_{y}|\zeta\rangle=sgn(B)\frac{iv_{F}}{2}(\lambda^{'}a_{n^{'}}b_{n}\delta_{n^{'},n+1}-\lambda a_{n}b_{n^{'}}\delta_{n^{'},n-1})\delta_{\nu,\nu^{'}}\delta_{k_{y}k_{y}^{'}},
\end{equation}
and, taking the integral over $k_{y}$, we have
\begin{equation}\label{eq10}
 \sigma_{xy}=-g_{s}g_{v}sgn(B)\frac{\hbar e^{2}v_{F}^{2}}{4\pi l_{B}^{2}}\sum_{\nu,n,n^{'},\lambda,\lambda^{'}}\frac{[f(\epsilon_{n}^{\lambda,\nu})-f(\epsilon_{n^{'}}^{\lambda^{'},\nu})]
 [a_{n^{'}}^{2}b_{n}^{2}\delta_{n^{'},n+1}-a_{n}^{2}b_{n^{'}}^{2}\delta_{n^{'},n-1}]}
{(\epsilon_{n}^{\lambda,\nu}-\epsilon_{n^{'}}^{\lambda^{'},\nu^{'}})^{2}}.
\end{equation}
This expression can be further simplification to arrive at
\begin{equation}\label{eq11}
\sigma_{xy}=-g_{s}g_{v}sgn(B)\frac{e^{2}}{h}\sum_{\nu,n}(n+\frac{1}{2})[f(\epsilon_{n}^{+,\nu})-
f(\epsilon_{n+1}^{+,\nu})+f(\epsilon_{n}^{-,\nu})-f(\epsilon_{n+1}^{-,\nu})].
\end{equation}
It is obvious that $\sigma_{xy}$ of the AAA-stacked TLG can be written in terms $\sigma_{xy}$ of SLG as
\begin{equation}\label{eq12}
\sigma_{xy}(\mu)=\sigma_{xy}^{-}+\sigma_{xy}^{0}+\sigma_{xy}^{+}=\sigma_{xy}^{SLG}(\mu+\sqrt{2}t_{\bot})+\sigma_{xy}^{SLG}(\mu)+\sigma_{xy}^{SLG}(\mu-\sqrt{2}t_{\bot}),
\end{equation}
namely the Hall conductivity of the AAA-stacked of TLG is three copies of that of SLG, but with different $\mu+\sqrt{2}t_{\bot}$, $\mu$ and $\mu-\sqrt{2}t_{\bot}$ chemical potential (denoted by $\sigma_{xy}^{-}$, $\sigma_{xy}^{0}$ and $\sigma_{xy}^{+}$ respectivley), where
\begin{equation}\label{eq13}
\sigma_{xy}^{SLG}=-g_{s}g_{v}sgn(B)\frac{e^{2}}{h}\sum_{n}(n+\frac{1}{2})[f(\epsilon_{n}^{+})-
f(\epsilon_{n+1}^{+})+f(\epsilon_{n}^{-})-f(\epsilon_{n+1}^{-})]
\end{equation}
with $\epsilon_{n}^{\pm}=\pm \hbar \omega_{c}\sqrt{n}$ being the Landau levels of SLG~\cite{Gusynin1,Tahir1}.

As shown in Ref. ~\cite{Gusynin1}, and as it is clear from Eq. \ref{eq13}, the Hall conductivity of SLG is a odd function of the chemical potential, $\sigma_{xy}^{SLG}(-\mu)=-\sigma_{xy}^{SLG}(\mu)$. We can write the Hall conductivity of the AAA-stacked TLG as $\sigma_{xy}(-\mu)=\sigma_{xy}^{SLG}(-[\mu-\sqrt{2}t_{\bot}])+\sigma_{xy}^{SLG}(-\mu)+\sigma_{xy}^{SLG}
(-[\mu+\sqrt{2}t_{\bot}])$. So, we arrive at the same result, $\sigma_{xy}(-\mu)=-\sigma_{xy}(\mu)$, for $\sigma_{xy}$ of the AAA-stacked TLG.

At zero temperature the Hall conductivity of SLG can be written as $\sigma_{xy}^{SLG}=-g_{s}g_{v}sgn(\mu)sgn(B) \frac{e^{2}}{h}(n+\frac{1}{2})$ where $n$ is the index number of the last occupied (first empty) Landau level for $\mu>0(\mu<0)$, and it is equal to the integer part of $\frac{\mu^{2}}{2\hbar|eB|v_{F}^{2}}$~\cite{Gusynin1}. Using this result we arrive at
\begin{equation}\label{eq14}
\sigma_{xy}=-g_{s}g_{v}sgn(B)\frac{e^{2}}{h}[sgn(\mu-\sqrt{2}t_{\bot})(n+\frac{1}{2})+sgn(\mu)(n^{'}+\frac{1}{2})+sgn(\mu+
\sqrt{2}t_{\bot})(n^{''}+\frac{1}{2})]
\end{equation}
for the Hall conductivity of the AAA-stacked TLG at zero temperature, where $n$, $n^{'}$ and $n^{''}$ are the index number of the last occupied (first empty) Landau level of the corresponding set of the Landau levels, which can be written as $n=[\frac{(\mu-\sqrt{2}t_{\bot})^{2}}{2\hbar|eB|v_{F}^{2}}]$, $n^{'}=[\frac{\mu^{2}}{2\hbar|eB|v_{F}^{2}}]$ and $n^{''}=[\frac{(\mu+\sqrt{2}t_{\bot})^{2}}{2\hbar|eB|v_{F}^{2}}]$, where $[x]$ is equal to the integer part of $x$.

\section{Results and Discussion}
\label{sec03}

In this section, utilizing the expression obtained in the previous section, we investigate $\mu$ and $1/B$ dependence of the Hall conductivity in AAA-stacked TLG. In our calculation we use $t=3~eV$, $t_{\bot}=0.2~eV$ and $0.142~nm$ values for the intra-, inter-layer hopping energies and intera-layer carbon bound length, respectively, and restrict our investigation to zero temperature.

Fig. \ref{Fig01} shows the Hall conductivity of the AAA-stacked TLG as a function of the inverse magnetic field for different values of the chemical potential. As it was proven above that $\sigma_{xy}(-\mu)=-\sigma_{xy}(\mu)$ and $\sigma_{xy}(-B)=-\sigma_{xy}(B)$, so we consider $\sigma_{xy}$ only for $\mu>0$ and $B<0$. In this figure, three points are worth mentioning. \textit{Firstly}, the Hall conductivity of the AAA-stacked TLG, similar to that of single layer graphene, exhibits a $\frac{4e^{2}}{h}$-step at $\frac{1}{B}=0$ for $\mu=0.100 eV$ and $\mu=0.200 eV$, while for other values it jumps by $\frac{12e^{2}}{h}$ on crossing $\frac{1}{B}=0$. In the following we will show that $\sigma_{xy}$ always exhibits $\frac{4e^{2}}{h}$ or $\frac{12e^{2}}{h}$-step at $\frac{1}{B}=0$, depending on whether the chemical potential is less or larger than $\sqrt{2}t_{\bot}$, respectively. \textit{Secondly}, for some values of the chemical potential $\sigma_{xy}$ shows, in addition to usual $\frac{4e^{2}}{h}$ steps, some $\frac{8e^{2}}{h}$ or $\frac{12e^{2}}{h}$-steps at $\frac{1}{B}\neq0$ which repeat periodically. \textit{Thirdly}, the Hall conductivity of the AAA-stacked TLG shows an unusual $1/B$ dependence (See the black curve), in which the Hall conductivity, depending on the value of the magnetic filed, may increase (or decrease) as the magnetic filed decreases gradually. This leads to an unusual quantization for the Hall conductivity. We show that this is a characteristic of multi-layer graphene structures with AA-stacking order, and this can also be observed in the AA-stacked BLG, although it is not mentioned in Ref. \cite{Hsu2}.

The first point could be explained as follows. We know that $\sigma_{xy}$ of AAA-stacked TLG is sum of three copies of $\sigma_{xy}$ of SLG with different chemical potentials, $\mu-\sqrt{2}t_{\bot}$, $\mu$ and $\mu+\sqrt{2}t_{\bot}$, coming from three decomposed sets of Landau levels (See Eq. \ref{eq04}). When $\mu<\sqrt{2}t_{\bot}$ two of these sets of Landau levels are electron-doped leading to positive Hall conductivity (for $B<0$) and the other one is hole-doped leading negative Hall conductivity. Furthermore, it is clear that the amount of the Hall conductivity at zero temperature is determined only by the index number of the last occupied (first empty) Landau level for electron(hole)-doped regime, $\sigma_{xy}^{SLG}=-sgn(B)sgn(\mu)\frac{4e^{2}}{h}(n+\frac{1}{2})$. For very strong magnetic field cases, the last occupied (first empty) Landau level of every electron-doped (hole-doped) set of the Landau levels is $n=0$ Landau level. Consequently, we arrive at $\sigma_{xy}(\frac{1}{B}\rightarrow 0^{-})=\frac{2e^{2}}{h}+\frac{2e^{2}}{h}-\frac{2e^{2}}{h}$ and $\sigma_{xy}(\frac{1}{B}\rightarrow 0^{+})=-\frac{2e^{2}}{h}-\frac{2e^{2}}{h}+\frac{2e^{2}}{h}$ which leads to $\frac{4e^{2}}{h}$-step at $\frac{1}{B}=0$ in the curves of $\sigma_{xy}(\frac{1}{B})$. If $\mu$ exceeds $\sqrt{2}t_{\bot}$ all sets of Landau levels become electron-doped, in which the last occupied Landau level of each set of the Landau levels (in the limit of $1/B\rightarrow 0$) are $n=0$ Landau level. So, we have $\sigma_{xy}(\frac{1}{B}\rightarrow 0^{-})=\frac{2e^{2}}{h}+\frac{2e^{2}}{h}+\frac{2e^{2}}{h}$ and $\sigma_{xy}(\frac{1}{B}\rightarrow 0^{+})=-\frac{2e^{2}}{h}-\frac{2e^{2}}{h}-\frac{2e^{2}}{h}$ which lead to $\frac{12e^{2}}{h}$-jump at $\frac{1}{B}=0$.

To explain the second point, first we focus on the Hall conductivity in SLG, and review the origin of the usual $\frac{4e^{2}}{h}$-steps change in the curve of $\sigma_{xy}^{SLG}(\frac{1}{B})$. Imagine an electron-doped graphene sheet in a vertical magnetic field. As the magnetic field increases gradually the energy distance between the adjacent Landau levels decreases and they pass through the Fermi energy one by one, and become filled, leading to $\frac{e^{2}}{h}$-steps increase in the Hall conductivity. Further, due to the spin and valley degeneracy, this effect takes place for both spin indices at both valleys simultaneously. Consequently, the Hall conductivity of SLG exhibits $\frac{4e^{2}}{h}$-steps at some $\frac{1}{B}\neq0$. For the AAA-stacked TLG, whose Landau levels are composed of three copies of SLG Landau levels, this condition can be fulfilled for two or three Landau levels from different sets of Landau levels at special values of the chemical potential and the magnetic field simultaneously. Namely, when the magnetic field increases gradually two or three Landau levels from different sets of Landau levels, at special values of the magnetic field, pass through the Fermi energy simultaneously. Consequently, the Hall conductivity of AAA-stacked TLG exhibits $\frac{8e^{2}}{h}$ or $\frac{12e^{2}}{h}$-steps change at $\frac{1}{B}\neq0$, and so some of the plateau don't appear in the curves of $\sigma_{xy}(\frac{1}{B})$ of the AA-stacked TLG. To be more accurate, these effects are observed when $\epsilon_{n}^{+,\nu}=\epsilon_{n^{'}}^{+,\nu^{'}}=\mu$ condition is satisfied for two of or all $\nu$ indices. Using this condition, we arrive at a general expression as
\begin{equation}\label{eq15}
\mu=\frac{\nu\sqrt{n^{'}}-\nu^{'}\sqrt{n}}{\sqrt{n^{'}}-\sqrt{n}}\sqrt{2}t_{\bot},~~~~~n^{'}>n=1,2,3,...
\end{equation}
for chemical potentials, at which $\frac{8e^{2}}{h}$-steps ($\frac{12e^{2}}{h}$-steps) appear in the curves of $\sigma_{xy}(\frac{1}{B})$ at especial values of the inverse magnetic field, when the condition of Eq. \ref{eq15} is fulfilled for two of $\nu$ (all $\nu$). For example, for the green curve of Fig. \ref{Fig01} which exhibits $\frac{8e^{2}}{h}$-steps the chemical potential is equal to $[+\sqrt{3}-(0)\sqrt{1}]\sqrt{2}t_{\bot}/[\sqrt{3}-\sqrt{1}]\approx0.386~eV$.  $\mu=\sqrt{6}t_{\bot}/[\sqrt{3}-1]$ satisfies the condition of Eq. \ref{eq15} only for $\nu=0$ and $-$, leading to $\frac{8e^{2}}{h}$-steps in the curve of the Hall conductivity. Furthermore, as it is clear, this condition is fulfilled not only for $(n^{'}=3,n=1)$ indices but also for other different pairs of $n$ and $n^{'}$, including $(6,2)$, $(9,3)$ and ... . If the condition is satisfied for the $n^{th}$ pair of the indices of the Landau levels the $n^{th}$ $\frac{8e^{2}}{h}$-step appears in the curves of $\sigma_{xy}(\frac{1}{B})$. Hence, the appearance of $\frac{8e^{2}}{h}$-steps in the curves of $\sigma_{xy}(\frac{1}{B})$, which accompanied with the absence of a plateau for each $\frac{8e^{2}}{h}$-step, repeat periodically. It is easy to show that between any two $\frac{8e^{2}}{h}$-steps the curves pass through $n^{'}+n-1$ plateau. Consequently, if $\mu>\sqrt{2}t_{\bot}$ ($\mu<\sqrt{2}t_{\bot}$) the $\bar{\nu}=4[m(n+n^{'}-1)+\frac{3}{2}]$ plateaus with $m=1, 2, 3, ...$ (the $\bar{\nu}=4[m(n+n^{'}-1)+\frac{1}{2}]$ plateaus with $m=1, 2, 3, ...$) disappear in the curves of the Hall conductivity of AAA-stacked TLG. Furthermore, as mentioned above, when the condition of Eq. \ref{eq15} is satisfied for all $\nu$ indices simultaneously, the Hall conductivity exhibits $\frac{12e^{2}}{h}$-steps at some $\frac{1}{B}\neq0$. For example see the purple curve of the Fig. \ref{Fig01}, in which $\mu=2\sqrt{2}t_{\bot}\approx0.566~eV$. For $\mu=2\sqrt{2}t_{\bot}$ the condition of Eq. \ref{eq15} is fulfilled for all $\nu$ at Landau levels with $(n,n^{'},n^{''})$ indices such as $=(1,4,9)$, $(2,8,18)$, $(3,12,27)$ and ... . The periodicity of the $\frac{12e^{2}}{h}$-steps and the absence of some plateau also takes place in this case. If the Hall conductivity exhibits $\frac{12e^{2}}{h}$-steps the $\bar{\nu}=4[m(n+n^{'}+n^{''}-1)+\frac{3}{2}]$ and $\bar{\nu}=4[m(n+n^{'}+n^{''}-2)+\frac{3}{2}]$ plateaus with $m=1, 2, 3, ...$ are absent. \textit{With these results in mind, we conclude that the Hall conductivity in AAA-stacked TLG, for $\mu>\sqrt{2}t_{\bot}$,  is quantized as
\begin{equation}\label{eq15}
\sigma_{xy}=\pm\frac{4e^{2}}{h}(m+\frac{3}{2})~~~~m=0, 1, 2, ... , 
\end{equation}
 provided $\mu\neq[(\nu\sqrt{n^{'}}-\nu^{'}\sqrt{n})/(\sqrt{n^{'}}-\sqrt{n})]\sqrt{2}t_{\bot}$. However, when $\mu=[(\nu\sqrt{n^{'}}-\nu^{'}\sqrt{n})/(\sqrt{n^{'}}-\sqrt{n})]\sqrt{2}t_{\bot}$ the Hall conductivity lacks $\bar{\nu}=4[m(n+n^{'}-1)+\frac{3}{2}]$  ( $\bar{\nu}=4[m(n+n^{'}+n^{''}-1)+\frac{3}{2}]$  and  $\bar{\nu}=4[m(n+n^{'}+n^{''}-2)+\frac{3}{2}]$ ) plateaus if the condition of Eq. \ref{eq15} is fulfilled for two of $\nu$ (all $\nu$) indices simultaneously.
}

To explore the origin of the unusual $1/B$ of the Hall conductivity of AAA-stacked TLG (the third point) in more detail, we have presented the plots of $\sigma_{xy}$ and its components ($\sigma_{xy}^{+}$, $\sigma_{xy}^{0}$ and $\sigma_{xy}^{-}$), as funtions of the inverse magnetc field for different values of the chemical potential in Fig. \ref{Fig02}. As it is clear from this figure, this unusual quantization of the Hall conductivity is due to competition between electron-doped and hole-doped Landau levels in the spectrum of the AAA-stacked TLG, which tend to change the Hall conductivity appositively. \textit{Consequently, Eq. \ref{eq14} can't be simplified any further to arrive at an universal quantization for the Hall conductivity which is valid for the $0<\mu<t_{\bot}$ interval}. It is evident that this unusual quantization can also be observed in other multi-layer graphene with AA stacking order whose Landau level spectrum at small chemical potential are composed of both electron-doped and hole-doped Landau levels. Figure \ref{Fig03} shows this effect in the AA-stacked bilayer graphene. One can see that, in contrast to what has been reported in Ref. \cite{Hsu2}, the $\sigma_{xy}=\pm\frac{4e^{2}}{h}n$ quantization is not valid for $0<\mu<t_{\bot}$ range entirely.

\section{Summary and conclusions}
\label{sec04}

In summary, we studied the integer quantum Hall effect in AAA-stacked TLG in the linear response regime. First we obtained the Landau level spectrum and the wavefunctions of AAA-stacked TLG. We found that the Landau level spectrum of AAA-stacked TLG are three copies of the Landau levels of SLG, two off them are shifted up and down by $\sqrt{2}t_{\bot}$ with respect to zero energy, where $t_{\bot}$ is intra-layer hopping energy. Then, using the Kubo formula we obtained an analytical expression for the Hall conductivity of AAA-stacked TLG. In the next section, utilizing this expression, we investigated the dependence of the Hall conductivity on the chemical potential and magnetic field at zero temperature. Interestingly, we found that quantization of the Hall conductivity of AAA-stacked TLG is severely dependent on the chemical potentials and the magnetic field. In particular, we showed that for $\mu>\sqrt{2}t_{\bot}$ the Hall conductivity is quantized as $\sigma_{xy}(1/B)=\pm\frac{4e^{2}}{h}(m+\frac{3}{2})$ where $m=0, 1, 2, ...$, provided $\mu\neq[\nu\sqrt{n^{'}}-\nu^{'}\sqrt{n})t_{\bot}/(\sqrt{n^{'}}-\sqrt{n})]\sqrt{2}t_{\bot}$ with $n^{'}>n=1, 2, 3, ...$, $\mu$ is the chemical potential. However, if $\mu=[\nu\sqrt{n^{'}}-\nu^{'}\sqrt{n})t_{\bot}/(\sqrt{n^{'}}-\sqrt{n})]\sqrt{2}t_{\bot}$  the Hall conductivity lacks some plateaus and at the same time exhibits, in addition to usual $\frac{4e^{2}}{h}$-steps, some $\frac{8e^{2}}{h}$-steps and $\frac{12e^{2}}{h}$-steps which repeat periodically. The origin of these effects, attributed to the special Landau level spectrum of the AAA-stacked TLG, was explained. As another notable result, we found that for $\mu<\sqrt{2}t_{\bot}$ the Hall conductivity of the AAA-stacked TLG shows an unusual $1/B$ dependence and don't obey a universal quantization condition which is valid for $0<\mu<\sqrt{2}t_{\bot}$ range entirely. The competition between the electron- and hole-doped Landau levels ofin the spectrum of the AAA-stacked TLG at small chemical potentials, which tend to change the Hall conductivity appositively, is attributed to be the origin of this effect. Finally, we showed that this is a characteristic of the multilayer graphene structures with AA stacking order whose Landau-levels spectrum at small chemical potentials are composed of both electron-doped and hole-doped Landau levels.

\textbf{Acknowledgment}
This work has been
supported by Farhangian university.

%

%
%
\newpage
\begin{figure}
\begin{center}
\includegraphics[width=16cm,angle=0]{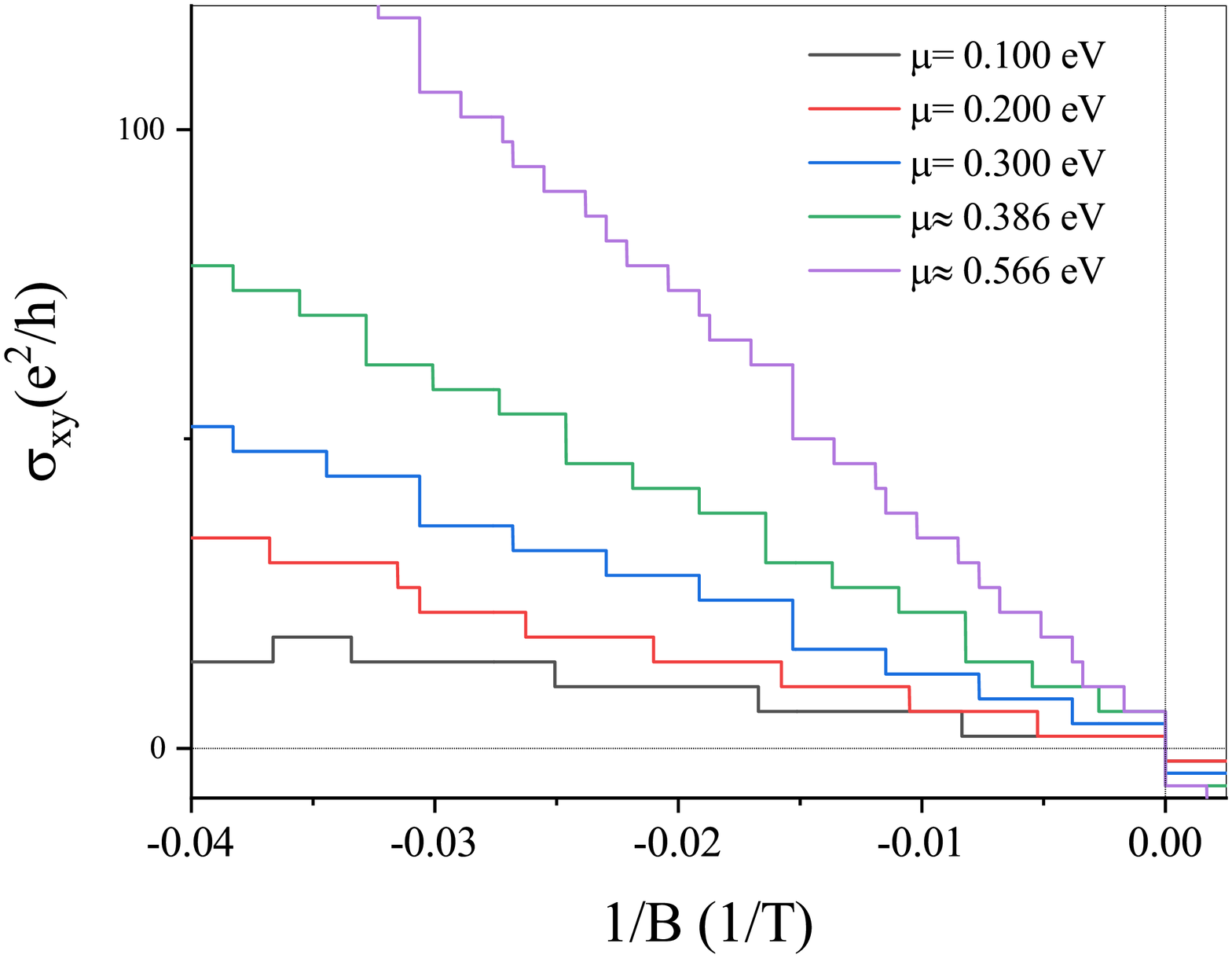}
\caption{The Hall conductivity of AAA-stacked TLG versus the inverse magnetic field for different values of the chemical potentials. The parameters $t=3~eV$, $t_{\bot}=0.2~eV$ and $0.142~nm$ are used for the intra-, inter-layer hopping energies and intera-layer carbon bound length, respectively.}\label{Fig01}
\end{center}
\end{figure}
\begin{figure}
\begin{center}
\includegraphics[width=16cm,angle=0]{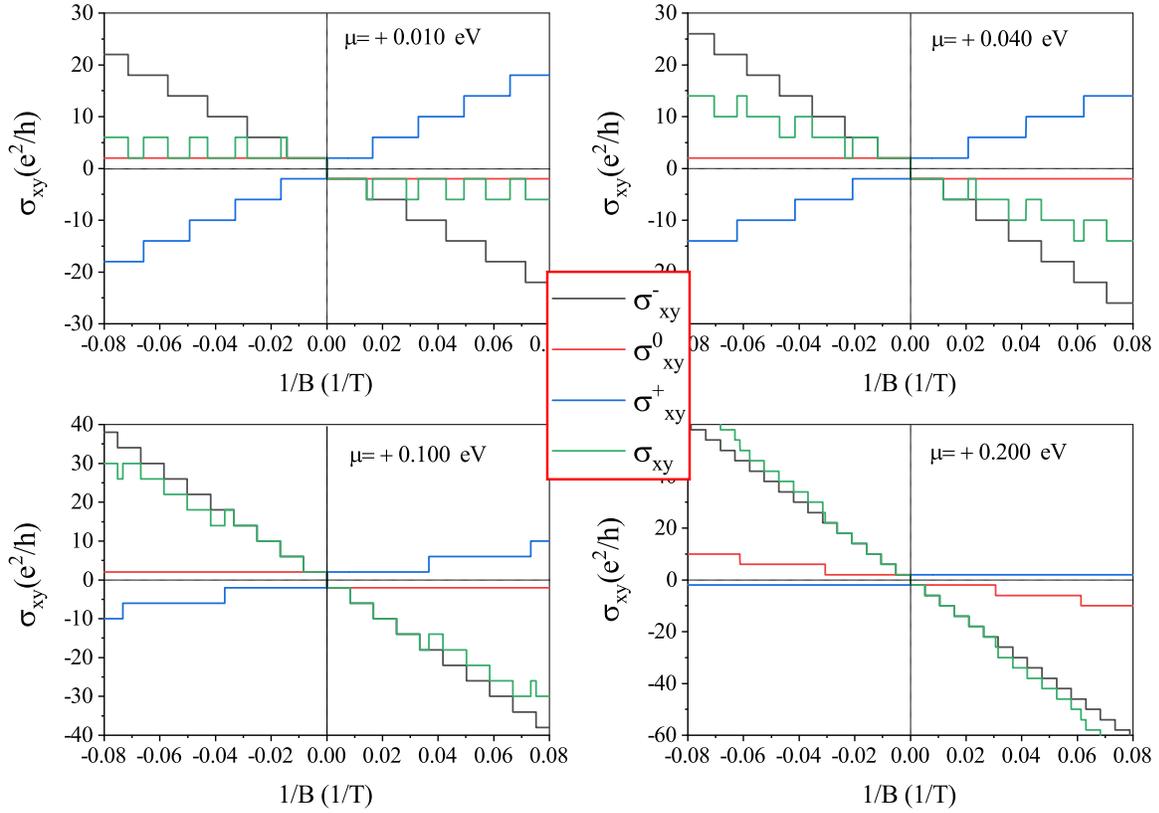}
\caption{The Hall conductivity of AAA-stacked TLG and its components as functions of the inverse magnetic field for different values of the chemical potentials, $\mu=0.01~eV$, $\mu=0.04~eV$, $\mu=0.1~eV$ and $\mu=0.2~eV$. The other parameters are $t=3~eV$, $t_{\bot}=0.2~eV$ and $0.142~nm$.}\label{Fig02}
\end{center}
\end{figure}
\begin{figure}
\begin{center}
\includegraphics[width=16cm,angle=0]{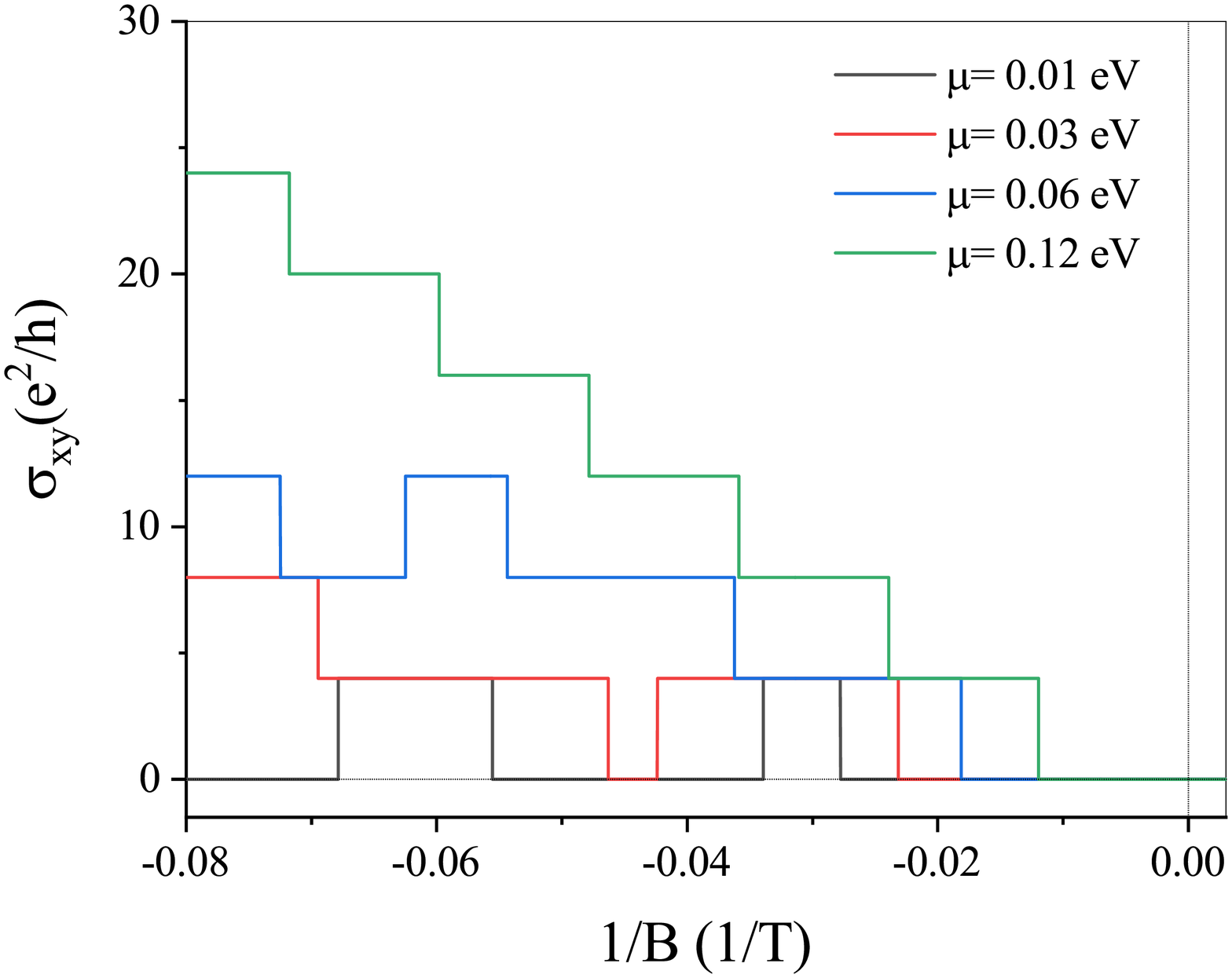}
\caption{The Hall conductivity of AA-stacked bilayer graphene versus the inverse magnetic field for different values of the chemical potentials. The other parameters are $t=3~eV$, $t_{\bot}=0.2~eV$ and $0.142~nm$.}\label{Fig03}
\end{center}
\end{figure}

\begin{thebibliography}{a}
%
\bibitem{Castro Neto1} A. H. Castro Neto, F. Guinea, N. M. R. Peres, K. S. Novoselov
and A. K. Geim, Rev. Mod. Phys. 81 (2009) 109-162.
\bibitem{Peres1} N. M. R. Peres, Rev. Mod. Phys. 82 (2010) 2673-2700.
%
\bibitem{Partoens1} B. Partoens, F. M. Peeters, Phys. Rev. B 75 (2007) 193402-193404.
\bibitem{Koshino1} M. Koshino, T. Ando, Phys. Rev. B 77 (2008) 115313-115323.
\bibitem{Nilsson1} J. Nilsson, A. H. Castro Neto, F. Guinea, N. M. R. Peres, Phys. Rev. B 78 (2008) 045405-045438.
\bibitem{Avetisyan1} A. A. Avetisyan, B. Partoens, F. M. Peeters, Phys. Rev. B 80 (2009) 195401-195411.
\bibitem{Koshino2} M. Koshino, Phys. Rev. B 81 (2010) 125304-125310.
\bibitem{Das Sarma1} S. Das Sarma, S. Adam, E. H. Hwang, E. Rossi, Rev. Mod. Phys. 83 (2011) 407-470.
\bibitem{Jung1} J. Jung, F. Zhang, Z. Qiao, A. H. MacDonald, Phys. Rev. B 84 (2011) 075418-075422.
\bibitem{de Andres1} P. L. de Andres, F. Guinea, M. I. Katsnelson, Phys. Rev. B 86 (2012) 144103-144107.
\bibitem{Munoz1} W. A. Munoz, L. Covaci, F. M. Peeters, Phys. Rev. B 88 (2013) 214502-214506.
%
\bibitem{Aoki1} M. Aoki and H. Amawashi, Solid State Commun. 142 (2007) 123-.
\bibitem{Andres1} P. L. de Andres, R. Ramlrez, and J. A. Verges, Phys. Rev. B 77 (2008) 045403-045403.
\bibitem{Lauffer1} P. Lauffer, K. V. Emtsev, R. Graupner, Th. Seyller, and
L. Ley, Phys. Rev. B 77 (2006) 155426-155435.
\bibitem{Liu1} Z. Liu, K. Suenaga, P. J. Harris, S. Iijima, Phys. Rev. Lett. 102 (2009) 015501-015504.
%
\bibitem{Borysiuk1} J. Borysiuk, J. Soltys, J. Piechota, Journal of Applied Physics 109 (2011) 093523-093527.
\bibitem{Ando1} T. Ando, Journal of Physics: Conference Series 302 (2011) 012015-012027.
%
\bibitem{Hsu1} Y. -F. Hsu, G. -Y. Guo, Phys. Rev. B 82 (2011) 165404-165411.
\bibitem{Prada1} E. Prada, P. San-Jose, L. Brey, H. Fertig, Solid State Commun. 151 (2011) 1065-1070.
\bibitem{Tabert1} C. J. Tabert, E. J. Nicol, Phys. Rev. B 84 (2012) 075439-075450.
\bibitem{Brey1} L. Brey, H. A. Fertig, Phys. Rev. B 87 (2013) 115411-115418.
\bibitem{Mohammadi1} Y. Mohammadi and B. Arghavani-Nia, Solid State Communic. 201 (2015) 76-81.
%
%
%
\bibitem{Klitzing1} K. v. Klitzing, G. Dorda and M. Pepper, Phys. Rev. Lett. 45 (1980) 494-497.
\bibitem{Novoselov1} K. S. Novoselov, A. K. Geim, S. V. Morozov, D. Jiang, M. I. Katsnelson, I. V. Grigorieva,
S. V. Dubonos and A. A. Firsov, Nature, 438 (2005) 197-200.
\bibitem{Gusynin1} V.P. Gusynin and S.G. Sharapov,  Phys. Rev. Lett. 95 (2005) 146801.
%
\bibitem{McCann1} E. McCann and V. I. Fal'ko, Phys. Rev. Lett. 96 (2006) 086805-086808
\bibitem{Koshino3} M. Koshino and T. Ando, Phys. Rev. B 77 (2008) 115313-115320.
\bibitem{Henriksen1} E. A. Henriksen, D. Nandi, and J. P. Eisenstein, Phys. Rev. X 2 (2012) 011004-011011.
\bibitem{Min1} H. Min and A. H. MacDonald, Phys. 77 (2008) 155416-155420.
\bibitem{Min2} H. Min and A. H. MacDonald, Prog. Thror. Phys. Supp. 176 (2008) 227-252.
\bibitem{Hsu2} Y. -F. Hsu and G. -Y. Guo, Phys. Rev. B 82 (2010) 165404.
%
%
%
\bibitem{Mohammadi2} Y. Mohammadi, R. Moradian and F. Shirzadi-Tabar, Solid State Communic. 193 (2014) 1-5.
\bibitem{Krstajic1} P. M. Krstajic and P. Vasilopoulos, Phys. Rev. B 86 (2012) 115432-115439.
\bibitem{Tahir1} M. Tahir and U. Schwingensvhogl, Sci. Rep. 3 (2012) 1075-1079.
%
%
%
\end{thebibliography}
\end{document}